\documentclass[11pt]{article}
\usepackage{moriond,epsfig}

\def\Journal#1#2#3#4{{#1} {\bf #2}, #3 (#4)}

\def\PLB{{\em Phys. Lett.}  B}
\def\PRL{\em Phys. Rev. Lett.}

\def\PRC{{\em Phys. Rev.} C}

\def\be{\begin{equation}}
\def\ee{\end{equation}}
\def\bea{\begin{eqnarray}}
\def\eea{\end{eqnarray}}

\newcommand{\iso}[2]{$^{#1}$#2}

\begin{document}
\vspace*{4cm}
\title{RESULTS FROM THE SALT PHASE OF SNO}

\author{ K. K. S. MIKNAITIS\\for the SNO Collaboration}
\address{Center for Experimental Nuclear Physics and Astrophysics,\\
  University of Washington, Seattle, WA 98195}

\maketitle\abstracts{
The Sudbury Neutrino Observatory (SNO) has recently completed an analysis of
data from the salt phase of the experiment, in which NaCl was added to the
heavy-water neutrino target to enhance sensitivity to solar
neutrinos. Results from the 391-day salt data set are summarized,
including the measured
solar neutrino fluxes, the electron energy spectrum from charged current
interactions, and the day-night neutrino flux asymmetries.  Constraints
on neutrino mixing parameters including the new measurements are also given.}

\section{Introduction}

The Sudbury Neutrino Observatory (SNO) is a heavy water Cherenkov detector
located 2092~m underground near Sudbury, Ontario.\cite{nim}  The
neutrino target consists of 1000 tonnes of ultra-pure D$_2$O housed in a
clear acrylic vessel 12 m in diameter.  SNO detects the neutrinos from
\iso{8}B decays in the sun through three reactions:
\begin{eqnarray}
\mbox{Charged Current Reaction (CC): \,\,\,\,\,\,\,\,\,   }\nu_e+\mbox{d}&\longrightarrow&\mbox{p}+\mbox{p}+e^{-}\nonumber\\
\mbox{Elastic Scattering Reaction (ES): \,\,\,\,\,\,\,\,   }\nu_x+e^{-}&\longrightarrow&\nu_x+e^{-}\nonumber\\
\mbox{Neutral Current Reaction (NC): \,\,\,\,\,\,\,\,\,
}\nu_x+\mbox{d}&\longrightarrow&\nu_x + \mbox{p}+\mbox{n}.
\end{eqnarray}
The charged current reaction is sensitive only to electron neutrinos,
while the neutral current reaction is sensitive to any active flavor,
$\nu_x$, $ x = e, \mu, \tau$.
The elastic scattering reaction has some sensitivity to non-electron
flavors, but is primarily sensitive to electron neutrinos.  The NC reaction
in SNO allows a measurement of the total flux of all active \iso{8}B
solar neutrinos, 
while a comparison of the neutrino fluxes measured through the NC and CC
reactions tests for solar neutrino flavor change.

Cherenkov radiation from the electrons produced in the CC and ES
reactions is detected in an array of approximately 9500
photomultiplier tubes (PMTs).  The neutrons liberated in the NC reaction must
be detected through a secondary capture reaction.  The SNO experiment
was designed to run in three phases, incorporating three distinct
neutron capture signatures to ensure a robust NC measurement.
 In each  phase, the neutron response features, systematic
 uncertainties, background characteristics, and analysis techniques differ. 
In the first phase of the experiment, neutrons were detected using the
6.25~MeV gamma signature from neutron capture on deuterium. 
Results from the first phase confirmed solar model predictions
of the \iso{8}B flux, demonstrated solar neutrino flavor change, and
contributed to determination of the underlying neutrino oscillation
parameters.~\cite{prl1}~\cite{prl2}~\cite{prl3}

In the second phase of the experiment, ultra-pure salt (NaCl) was
dissolved in the D$_2$O, to a concentration of ($0.196\pm0.002)\%$ by
weight.   The higher cross section for neutron capture on
\iso{35}Cl and the higher energy (8.6~MeV) released in the reaction
improve SNO's efficiency for detecting NC neutrons.  In addition, the multiple
gammas produced in the \iso{35}Cl capture reaction result in a more
isotropic distribution of Cherenkov light relative to the distribution
of light in single-electron events.  Using the 
isotropy of light produced in a neutrino event, the CC and NC
signals can be statistically separated without requiring constraints
on the solar neutrino energy spectrum.  First
neutrino flux results for 254 days of salt data 
were published in 2004.\cite{prl4}  Results for the 391-day salt data set
have recently been reported \cite{nsp} and are summarized in this
article.    

Finally, the third phase of the SNO experiment began in the fall of 2004,
following the deployment of 36 strings of \iso{3}He proportional
counters and 4 strings 
of \iso{4}He proportional counters
inside the heavy water volume.  Neutron capture in the \iso{3}He
counters allows an event-by-event determination of the NC
rate, completely decoupled from the PMT detection of the CC and ES signals.  

\section{Data Analysis in the Salt Phase}

\begin{figure}[h]
\begin{center}
$\begin{array}{c@{\hspace{0.2cm}}c}
\epsfxsize=3.0in \epsfysize = 2.2in
\epsffile{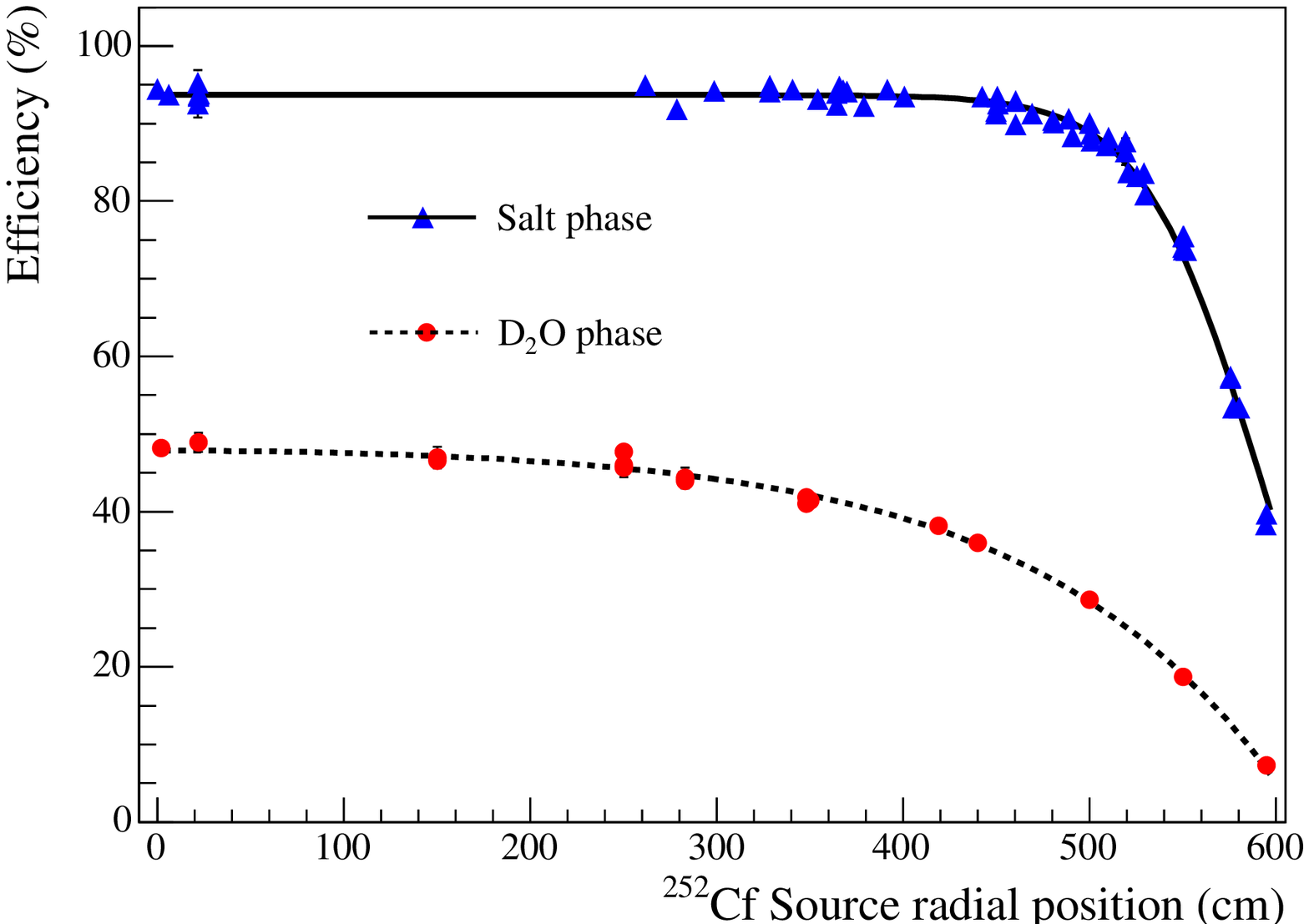} &
	\epsfxsize=3.0in 
	\epsffile{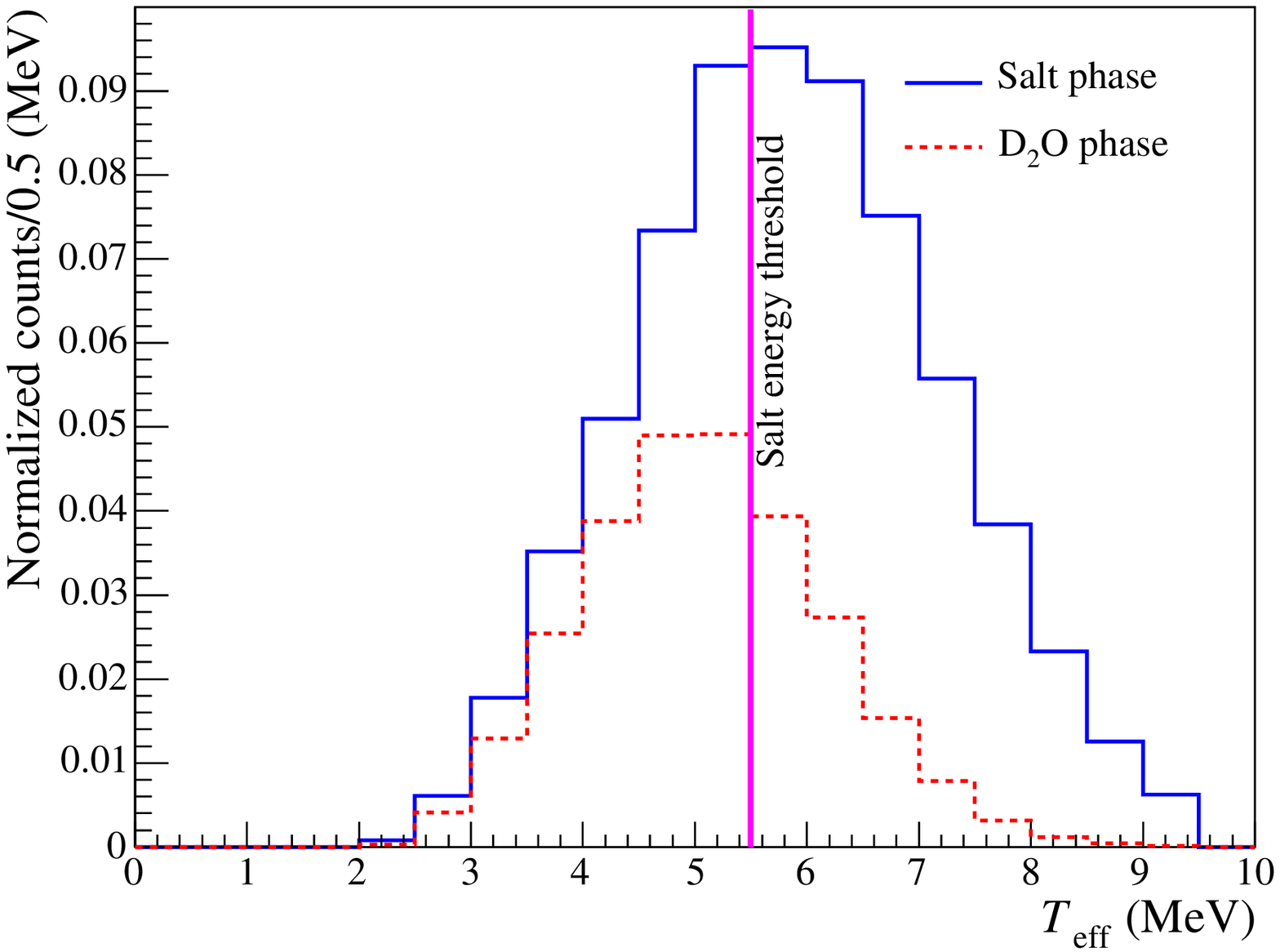}\\ [0.1cm]
\mbox{\bf (a)} & \mbox{\bf (b)}
\end{array}$
\end{center}
\caption{(a) Comparison of the neutron capture efficiency in the salt
  and D$_2$O phases.  The capture efficiency is shown for neutrons
from a \iso{252}Cf fission source of known strength, as a function of
the radius of the source in the detector.  (b)
  Comparison of the energy spectrum of neutrons events in the salt
  phase to the D$_2$O phase, as calculated by Monte Carlo. }
\label{fig:ncapt}
\end{figure}

The effect of the added salt on SNO's neutron detection efficiency is
illustrated in figure \ref{fig:ncapt}(a), which shows the neutron
capture efficiency for a \iso{252}Cf calibration source as a function of
the radial position of the source in the detector volume.  Accounting for the
fiducial volume and energy cuts used in the solar neutrino analysis,
the neutron capture efficiency in the salt phase is
40.7\%, compared to 14.4\% in the pure-D$_2$O phase.  
The higher energy of neutron capture events in the salt phase also
boosts the NC signal further above the analysis energy threshold.  A
Monte Carlo comparison of the energy spectrum for neutrons in the salt
and pure-D$_2$O phases is shown in figure \ref{fig:ncapt}(b).

A fiducial volume cut of 550~cm and an analysis energy threshold of
$T_{eff} > 5.5$~MeV are used to reduce the number of background events in the final
salt event sample, where $T_{eff}$ is the effective electron kinetic
energy for the event, $T_{eff} = E - 0.511$~MeV.  The final data set
consists of 4722 events recorded over 391.4 days of detector live time between July 2001 and
August 2003.  This data set contains
some remaining backgrounds as well as neutrino
interaction events.  The dominant background in SNO is
neutrons from the photodisintegration of deuterium by gammas with
energies above 2.2~MeV.  Radioactive contaminants from the U and
Th decay chains are the primary source of these gammas, which can
originate within the D$_2$O or in the acrylic vessel and external regions
of the detector.  The total contribution of neutron backgrounds from
inside the D$_2$O region is determined, using a variety of in-situ and
ex-situ techniques, to be $125.1^{+37.3}_{-32.0}$ events.
Additionally, an estimated $3.2^{+4.6}_{-4.4}$ events in the salt data
set are gamma rays from the products of atmospheric neutrino
interactions.  These ``internal'' neutron and gamma backgrounds are held
fixed in the statistical signal extraction process that is used to
determine the CC, NC, and ES reaction rates.  Neutrons and gamma rays
produced outside the D$_2$O region can also propagate into the
fiducial volume.  Neutron backgrounds due to these ``external''
sources will have a characteristic radial profile that can be used to
distinguish them from signal neutrons.

To extract the numbers of CC, NC, ES, and external neutron (EN)
events in the data set, event distributions in several variables are
fit to characteristic signal distributions derived from
Monte Carlo calculations.  The event variables used are the effective
electron kinetic energy $T_{eff}$, the radius
of the reconstructed event vertex $\rho = (r/600.5\mbox{cm})^3$, the
cosine of the angle between the reconstructed event direction and a
vector from the sun $\cos\theta_{\odot}$, and a
parameter characterizing the light isotropy in the event, $\beta_{14}$.
The $\beta_{14}$ parameter is based on
Legendre polynomials in terms of the angle between each pair
of hit PMTs relative to the event vertex.  Events with a
more isotropic distribution of light have smaller values of $\beta_{14}$.  
\begin{figure}[h]
\begin{center}
$\begin{array}{c@{\hspace{0.2cm}}c}
\epsfxsize=2.4in 
\epsffile{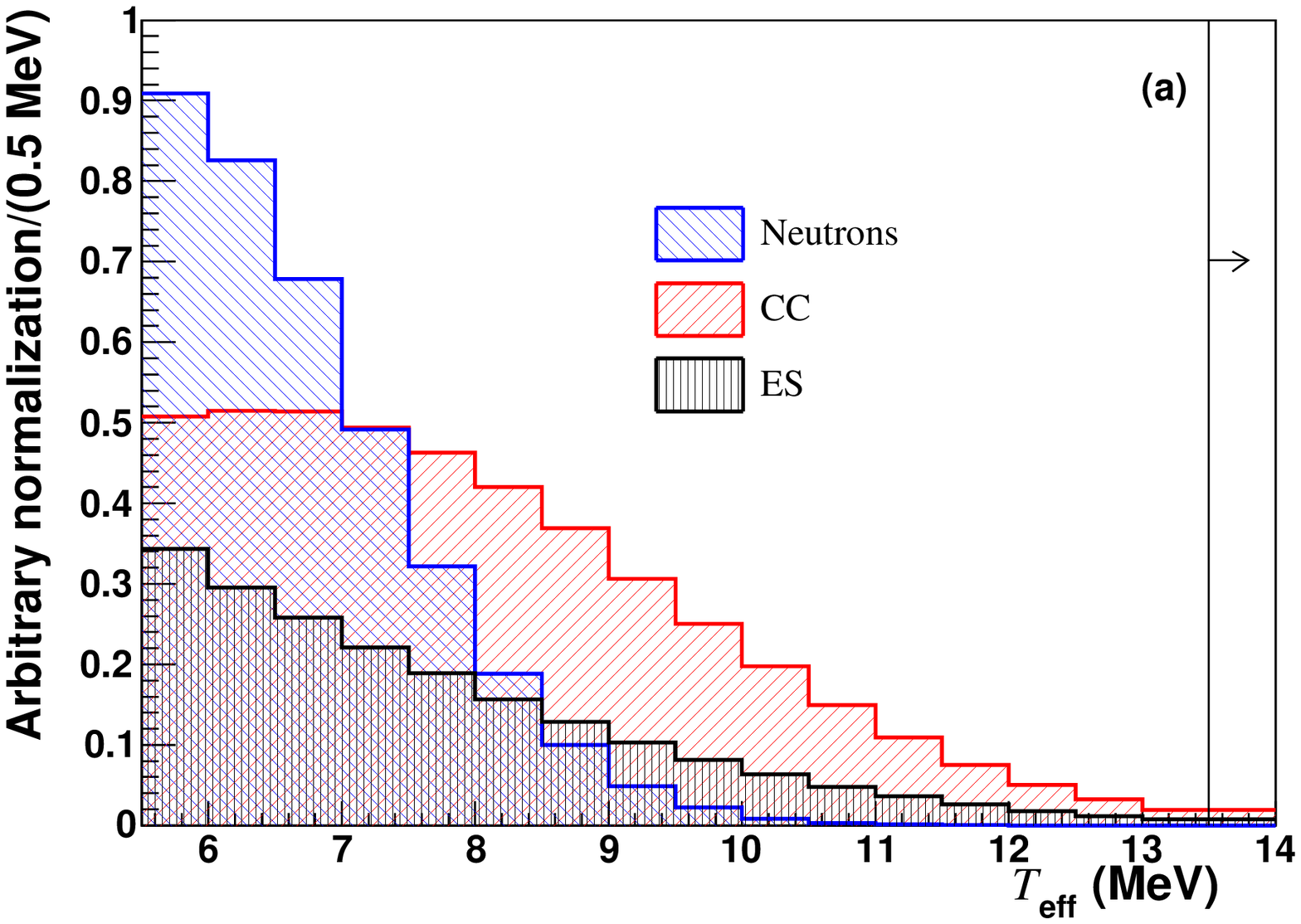} &
	\epsfxsize=2.4in 
	\epsffile{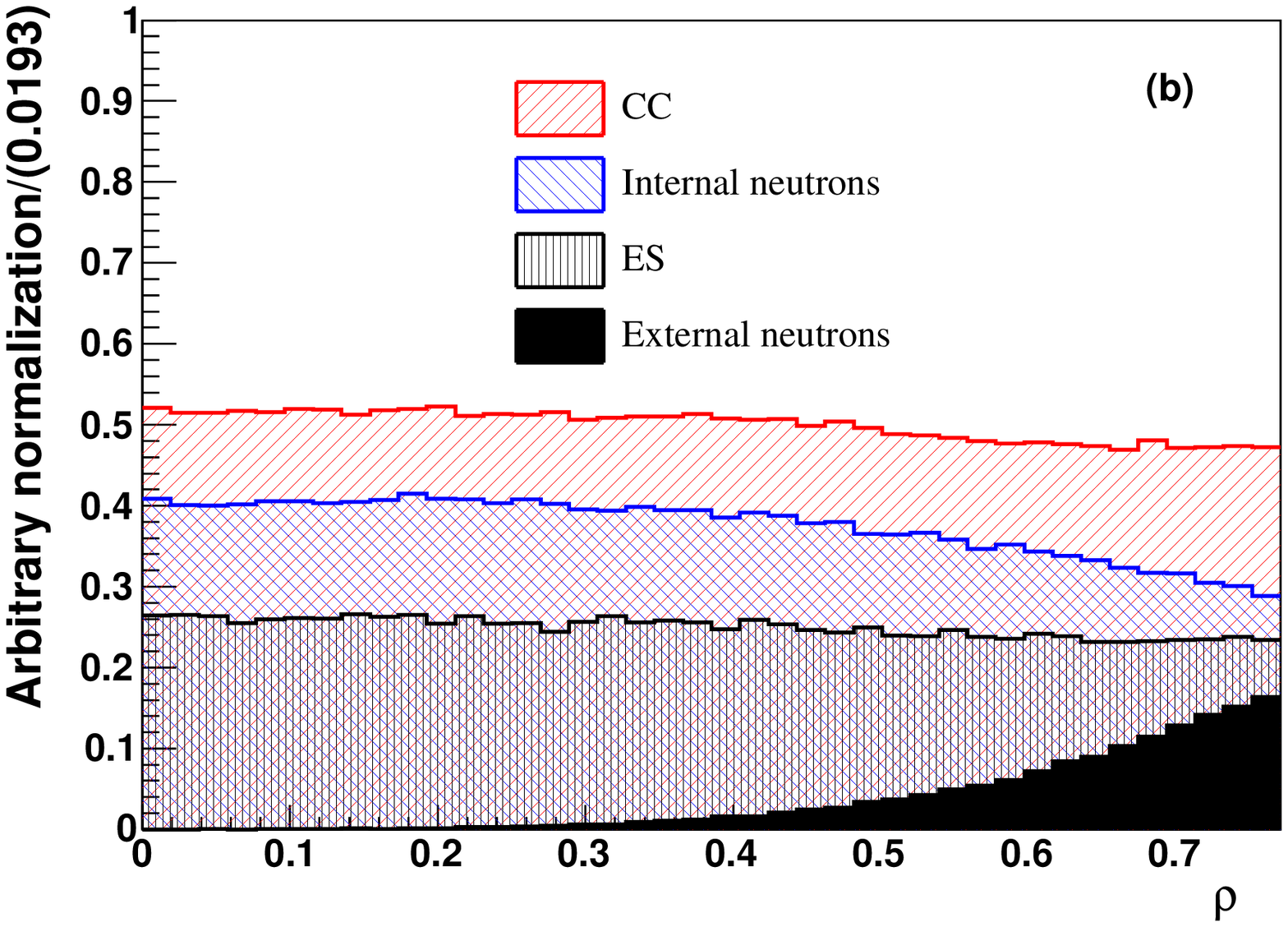} \\ [0.2cm]
\epsfxsize=2.4in 
\epsffile{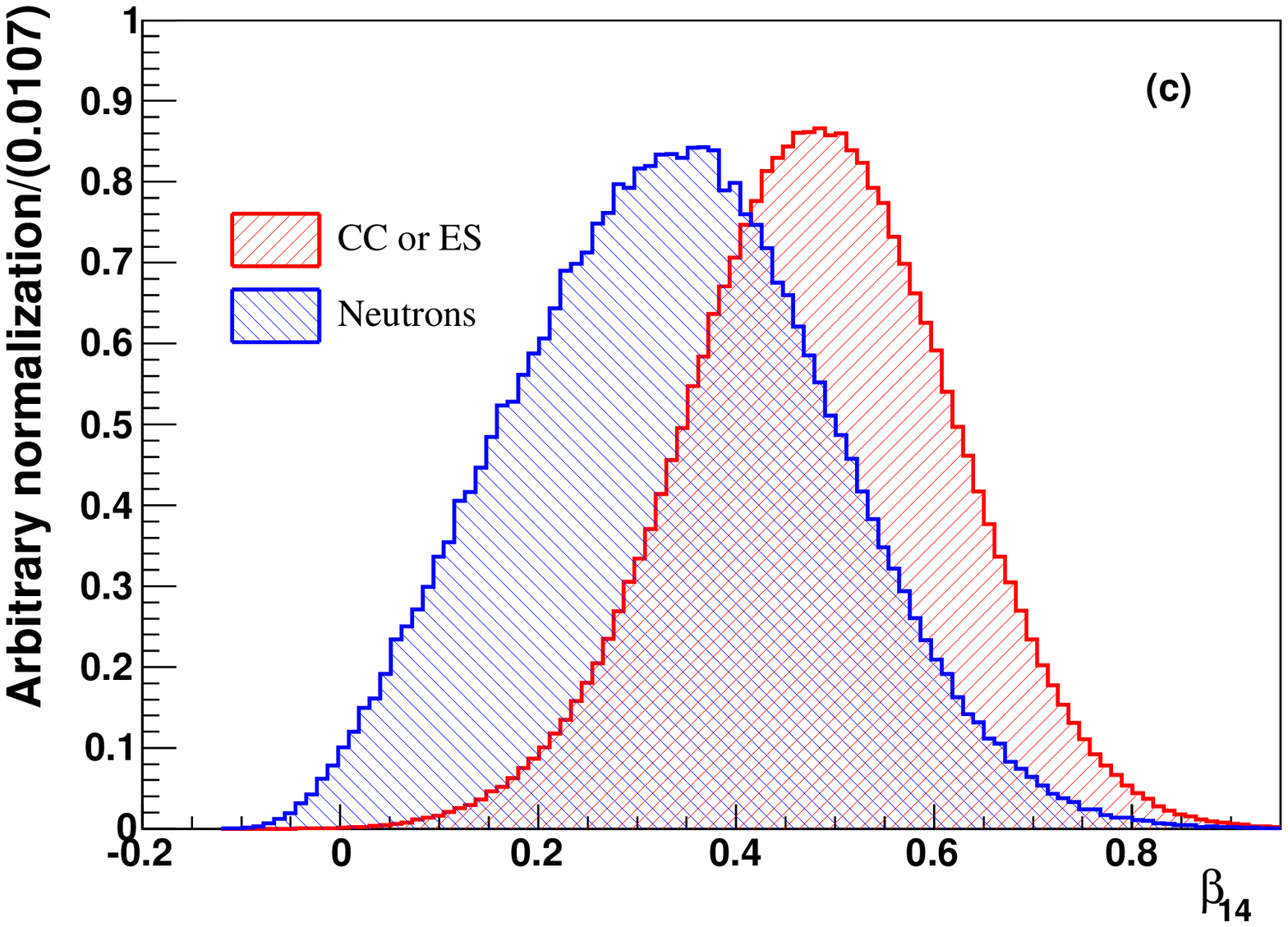} &
	\epsfxsize=2.4in 
	\epsffile{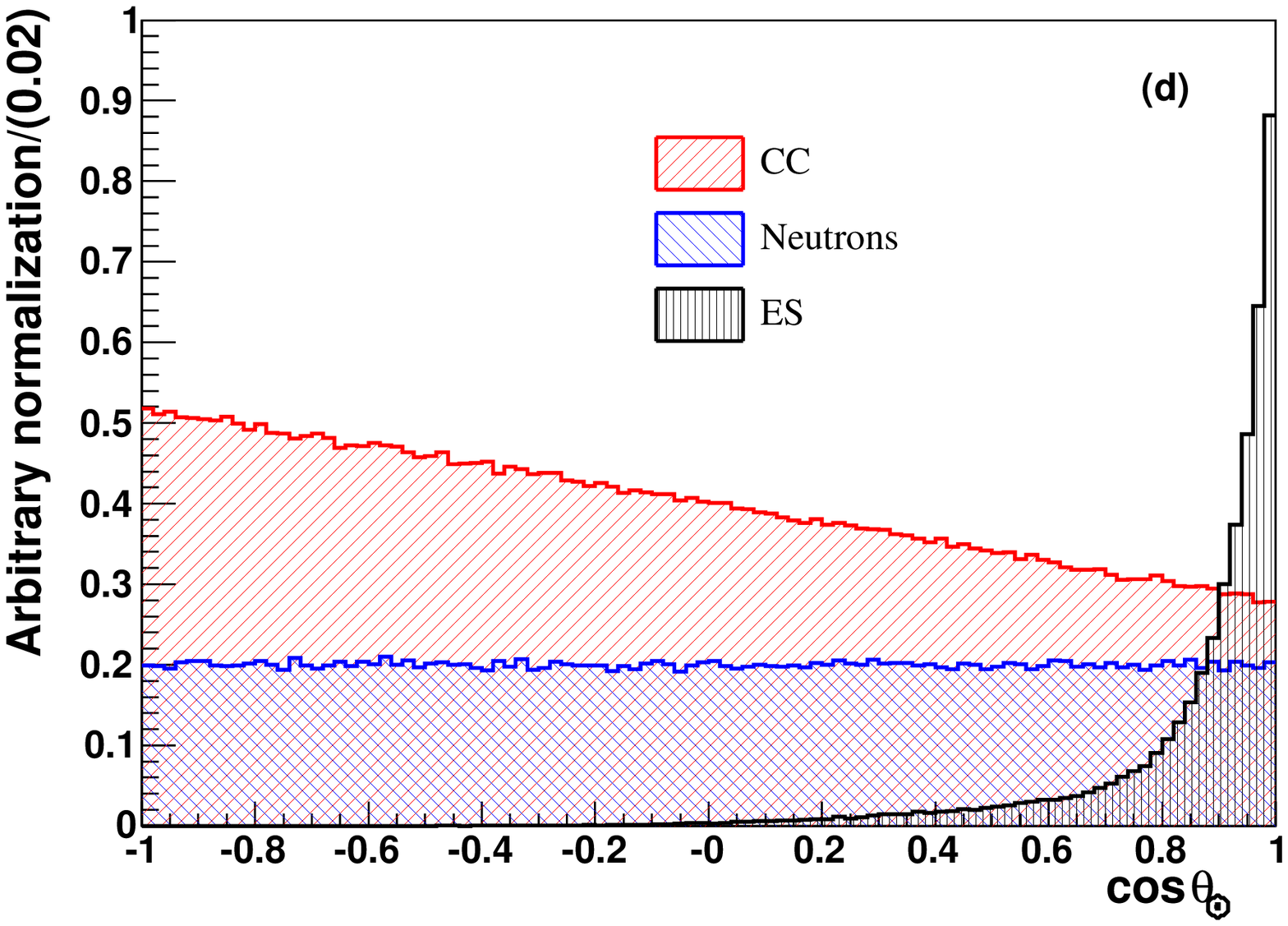}
\end{array}$
\end{center}
\caption{Monte Carlo probability density functions used in signal
  extraction, for (a) energy, (b) radius of the event vertex, (c)
  isotropy, and (d) direction relative to the sun.  Normalizations are
  arbitrary.  Where external neutron backgrounds are indistinguishable
  from internal neutrons 
(including those produced by the NC signal), the distribution is
simply labeled ``neutrons''.  The CC energy shape shown in (a)
corresponds to an undistorted \iso{8}B neutrino spectrum.}
\label{fig:pdfs}
\end{figure}

Detailed Monte Carlo simulations are used to generate probability
density functions (PDFs) characterizing signal distributions in these
variables.  These PDFs are shown in figure \ref{fig:pdfs}.
In a statistical fit, the isotropy distributions
provide a powerful tool for distinguishing neutrons from CC and ES events,
the direction distributions help to identify the ES contribution,
and the radial distributions separate the external neutron background.

The energy distribution shown for the CC signal
assumes an undistorted \iso{8}B neutrino spectrum.  To make
measurements that are independent of model assumptions about the
energy spectrum, we can extract the CC and ES signals separately in
each energy bin.  In such an ``energy unconstrained'' analysis, the
extracted CC spectrum can be used to test models of neutrino
oscillation.  This model-independent signal extraction is possible
because of the power of the $\beta_{14}$ distribution for separating
event classes 
in salt.  In the D$_2$O phase, the CC and ES spectra had to be
constrained to break statistical correlations between the signals.  

\section{Systematic Uncertainties}

Systematic uncertainties in detector response are evaluated 
through comparisons of Monte Carlo simulations and calibration data.
The primary calibration sources used to study systematic uncertainties
are a \iso{16}N 6.13~MeV gamma-ray source and a \iso{252}Cf fission
neutron source.  The \iso{16}N source is used to study energy
response, event reconstruction performance, and 
detector stability over time.  The \iso{252}Cf source is used to
evaluate neutron response characteristics.  
Systematic uncertainties are
propagated by perturbing the PDFs according to the estimated 1$\sigma$
variation in each response parameter, and then repeating the signal
extraction process. 

The dominant systematic uncertainties on the CC and NC extracted
fluxes in the energy-unconstrained analysis are due to uncertainty in
the $\beta_{14}$ parameter.  Uncertainties of less than a percent in the
mean isotropy values translate to uncertainties of around 4\% in the CC and
NC fluxes.  The energy scale uncertainty in the salt phase is
estimated to be 1.15\%, which contributes an uncertainty of around
3.5\% in the NC flux, but has a smaller effect on the CC and ES
fluxes.  An uncertainty of 1\% in radial reconstruction accuracy
is also one of the larger 
contributions to the overall systematic error, resulting in a $\sim3$\%
uncertainty in each flux.  The ES flux uncertainty is dominated by
a 5\% systematic uncertainty due to uncertainty in angular resolution.
A much more detailed discussion of calibrations and systematic
uncertainties can be found in the recent salt phase publication.\cite{nsp}

\section{Solar Neutrino Flux Results from an Energy-Unconstrained Analysis}

From an energy-unconstrained extended maximum likelihood fit to extract
the contributions of each signal to the data set, the
number of NC events is $2010\pm85$, the number of
CC events is $2176\pm78$, and the number of ES events is
$279\pm26$.  The external neutron background is $128\pm42$ events.  
Accounting for acceptance factors and detector live time, we can
convert these extracted event numbers into equivalent fluxes of
\iso{8}B solar neutrinos, in units of $10^6\,\mbox{cm}^{-2}\mbox{s}^{-1}$, 
\begin{eqnarray}
\phi_{\mbox{\small{CC}}}& =& 1.68^{+0.06}_{-0.06}\mbox{(stat.)}^{+0.08}_{-0.09}\mbox{(syst.)}\nonumber\\
\phi_{\mbox{\small{ES}}}& =& 2.35^{+0.22}_{-0.22}\mbox{(stat.)}^{+0.15}_{-0.15}\mbox{(syst.)}\nonumber\\
\phi_{\mbox{\small{NC}}}& =& 4.94^{+0.21}_{-0.21}\mbox{(stat.)}^{+0.38}_{-0.34}\mbox{(syst.)}.
\end{eqnarray}

\begin{figure}
\begin{center}
\psfig{figure=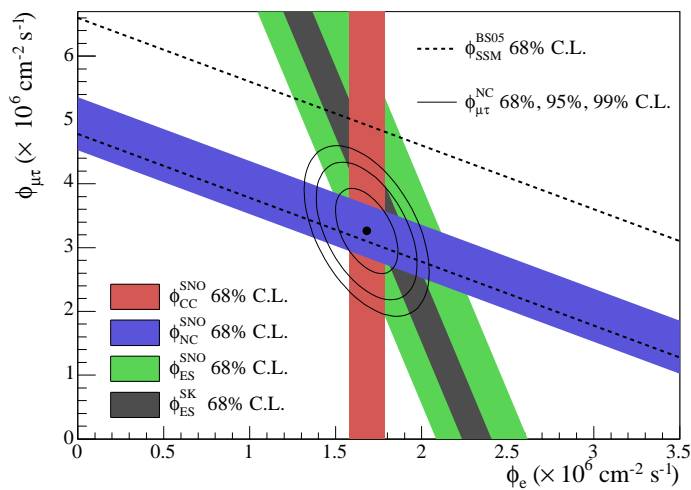,height=2.8in}
\caption{\label{fig:pnt} Flux of $\mu$ and $\tau$ neutrinos versus flux of
  electron neutrinos.  CC, NC, and ES flux measurements are indicated
  by the filled bands.  The total \iso{8}B solar neutrino flux
  predicted by the Standard Solar Model is shown as dashed lines,
  parallel to the NC measurement.  The narrow band parallel to the SNO
  ES measurement corresponds to the Super-Kamiokande elastic
  scattering result.  The best-fit point is determined using only the SNO data.}\end{center}
\end{figure}

The flavor
composition of \iso{8}B solar neutrinos detected by SNO is
summarized in figure \ref{fig:pnt}, which also shows comparisons to
the ES flux measured by the Super-Kamiokande experiment~\cite{sk} and
the Standard Solar Model predicted total flux.~\cite{ssm} The best-fit
point indicates the appearance of non-electron neutrino flavors in the
solar neutrino flux.

\section{Charged Current Spectrum and Day-Night Asymmetries}

The favored model for explaining solar neutrino flavor
change invokes matter-enhanced neutrino oscillation in the solar
interior, through the so-called Mikheyev-Smirnov-Wolfenstein (MSW) effect.  
In addition to predicting solar neutrino flavor change, the MSW effect
has two other predictions that are potentially testable in SNO.
Matter effects in the sun 
could distort the energy spectrum of solar neutrinos, and additional
matter effects in the earth could affect the flavor composition of
neutrinos that pass through terrestrial material before reaching a
detector.  The energy-unconstrained analysis of SNO's salt phase data allows an
extraction of the CC electron energy spectrum, which can be
used to test for spectrum distortions.  Matter effects in the earth
can be tested by looking for day-night asymmetries in the rate of
charged current interactions.  

For both the spectrum and day-night analyses, a number of differential
systematic uncertainties were evaluated.  Calibration sources were
used to study energy-dependent systematic effects. 
In-situ techniques using low-energy backgrounds and secondary products 
from cosmic-ray muon interactions were used to study diurnal variations
for the day-night analysis.   
\begin{figure}
\begin{center}
\psfig{figure=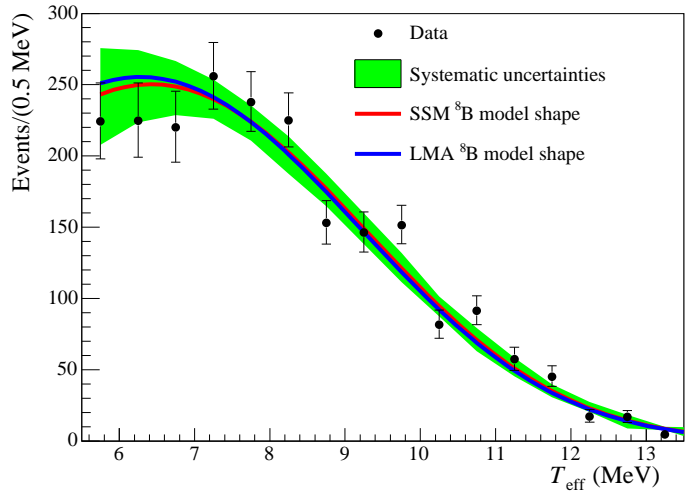,height=2.8in}
\caption{Extracted electron energy spectrum from the CC
  reaction, with statistical uncertainties. Systematic uncertainties
  are shown with 
  respect to the predicted spectrum for an undistorted \iso{8}B solar
  neutrino spectrum.  The spectrum shape for the best-fit MSW point
  (in the ``LMA'' region of the parameter space) is also shown.
\label{fig:spec}}
\end{center}
\end{figure}

The extracted CC energy spectrum is shown in figure \ref{fig:spec},
with statistical uncertainties.  Systematic
uncertainties are shown with respect to the prediction assuming an
undistorted \iso{8}B solar neutrino spectrum.  The measured spectrum
is consistent with no distortions, and is also consistent
with the spectrum predicted by the best-fit MSW model, corresponding
to the so-called Large Mixing Angle (LMA) region of the neutrino
oscillation parameter space.  

\begin{figure}
\begin{center}
\psfig{figure=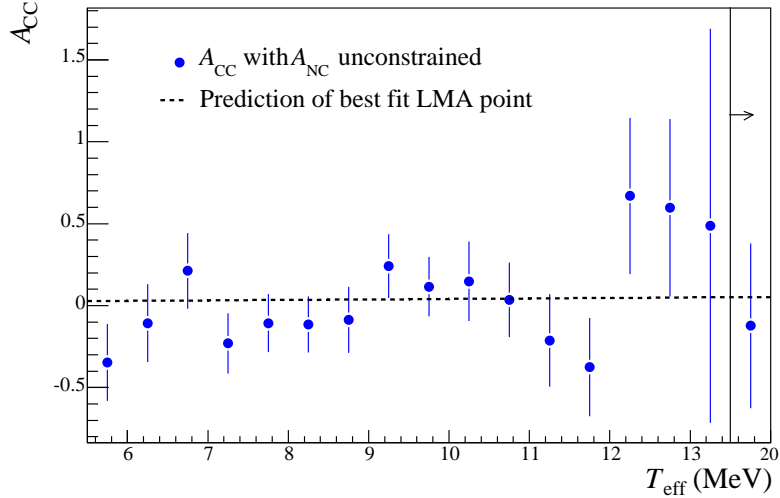,height=2.8in}
\caption{Charged current day-night asymmetry as a function of energy,
  with statistical uncertainties only.  The final bin extends from
  13.5~MeV to 20~MeV.  The dashed line shows the prediction from the
  previous best-fit point in the MSW parameter space.
\label{fig:acc_vs_e}}
\end{center}
\end{figure}

To search for day-night effects, we
construct asymmetry parameters in terms of the measured day and night fluxes,
$A_\alpha=
2(\Phi_{\alpha,N}-\Phi_{\alpha,D})/(\Phi_{\alpha,N}+\Phi_{\alpha,D})$,
where $\alpha$ = CC, NC, ES.
The asymmetries determined in an energy-unconstrained analysis of the
salt phase data are
\begin{eqnarray}
A_{{CC}} &=&
-0.056\pm0.074(\mbox{stat.})\pm0.053(\mbox{syst.})\nonumber\\
A_{{NC}} &=&
0.042\pm0.086(\mbox{stat.})\pm0.072(\mbox{syst.})\nonumber\\
A_{{ES}} &=& 0.146\pm0.198(\mbox{stat.})\pm0.033(\mbox{syst.}).
\end{eqnarray}
All asymmetries are consistent with zero. 
The extracted day and night CC spectra from this analysis 
can be used to construct the CC asymmetry as a function of energy,
shown in figure \ref{fig:acc_vs_e}.  Neutrino oscillation parameters
in the LMA region predict very small
day-night asymmetries, and a comparison to the LMA predicted asymmetry
is shown in the figure.  

Since an asymmetry in the NC rate is not expected for standard
neutrino oscillations, a statistically significant NC asymmetry would
be evidence for sterile neutrinos or exotic physics.  Within the
standard neutrino oscillation picture, we can constrain the neutral
current asymmetry to be zero to reduce the statistical correlations in
the extraction of the day-night asymmetries.  With this
additional constraint, the asymmetries become
\begin{eqnarray}
A_{{CC}} &=&
-0.037\pm0.063(\mbox{stat.})\pm0.032(\mbox{syst.})\nonumber\\
A_{{ES}} &=& 0.153\pm0.198(\mbox{stat.})\pm0.030(\mbox{syst.}).
\end{eqnarray}
Repeating this analysis with the assumption of an undistorted
\iso{8}B neutrino energy spectrum further reduces the uncertainties,
and allows the salt phase charged-current asymmetry to be combined with the
energy-constrained asymmetry results from the first phase of SNO.  The
combined day-night asymmetry from both phases is $A_{\mbox{\tiny{salt+D$_2$O}}} =
0.037\pm0.040$(stat.+syst.).  The day-night asymmetry results from SNO 
are consistent with no asymmetry, and also with the best-fit MSW
model, which predicts an asymmetry of $\sim3.5$\%.
 
\section{MSW Parameter Constraints}

\begin{figure}
\begin{center}
\psfig{figure=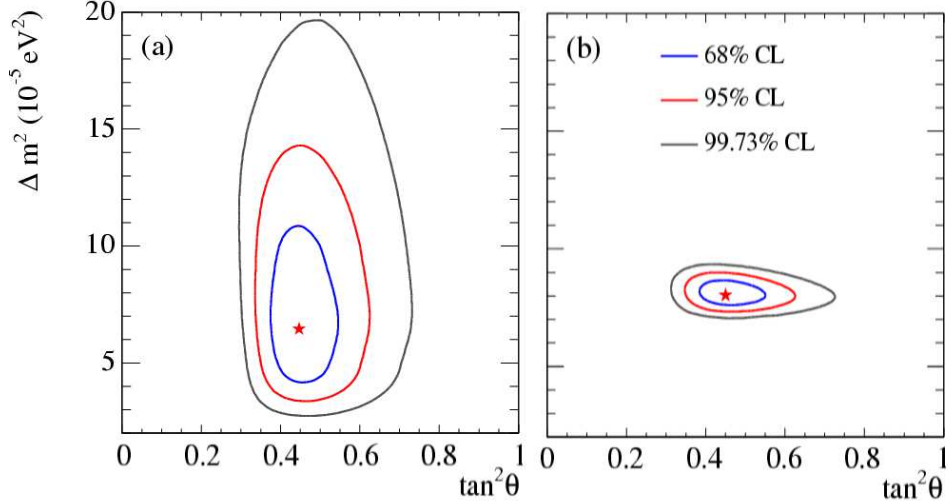,height=2.8in}
\caption{(a) Parameter constraints from a global neutrino oscillation analysis including fluxes and
  day and night energy spectra from SNO's salt and D$_2$O phases, as
  well as rate measurements from the Chlorine, SAGE, and Gallex/GNO
  experiments, and zenith spectra from Super-Kamiokande.  (b)
  Constraints from a global
  analysis including the KamLAND 766 ton-year
  results as well.  Best fit points are marked with stars.
\label{fig:msw}}
\end{center}
\end{figure}

The salt phase results for the fluxes, spectra, and day-night
asymmetries can be combined with SNO's previous results and the
results of other solar neutrino experiments to
produce constraints on the fundamental neutrino parameters in the MSW
model.  Figure \ref{fig:msw}(a) shows the results of a global
chi-squared analysis in a two-neutrino oscillation framework,
including SNO's salt phase and D$_2$O phase data as well as results
from Chlorine~\cite{cl} and Gallium~\cite{ga1}~\cite{ga2} experiments and the 
Super-Kamiokande experiment~\cite{sk}.  The best-fit point is $\Delta m^2 =
(6.5^{+4.4}_{-2.3})\times10^{-5}\mbox{eV}^2$, $\tan^2\theta =
0.45^{+0.09}_{-0.08}$.  Including
the results from the KamLAND experiment~\cite{kl1}~\cite{kl2} gives the contours shown in
figure \ref{fig:msw}(b).  The best fit point for the global analysis
including the KamLAND data is $\Delta m^2 =
(8.0^{+0.6}_{-0.4})\times10^{-5}\mbox{eV}^2$, $\tan^2\theta = 0.45^{+0.09}_{-0.07}$. 

\section{Conclusions}
Solar neutrino results from the salt phase of the SNO experiment have been
summarized in this paper.  These results include 
measurements of the flux of \iso{8}B solar 
electron neutrinos through the CC reaction, and
of the flux of \iso{8}B solar neutrinos of 
all active flavors through the NC reaction.  Use of the isotropy
parameter in the salt phase
allows NC and CC events to be statistically
separated without any assumptions about the underlying neutrino
energy spectrum.
The new flux
results confirm and improve previous results, demonstrate solar
neutrino flavor change, and contribute to evidence for solar neutrino
oscillations.  

Salt phase results also include the first 
presentation of the extracted charged current electron energy spectrum
with statistical and systematic uncertainties fully evaluated.  This
spectrum is consistent with the spectrum predicted for an undistorted
\iso{8}B neutrino spectrum, and is also consistent with the
spectrum predicted by the best-fit MSW model.
Day-night asymmetries have been constructed to test for possible MSW
effects in the earth.  Measured asymmetries are consistent with zero,
and also with the predictions for the best-fit MSW model.  A global
analysis of the salt phase results along with other solar and reactor
neutrino measurements has been performed, giving best-fit values of
$\Delta m^2 = 
(8.0^{+0.6}_{-0.4})\times10^{-5}\mbox{eV}^2$,$\tan^2\theta =
0.45^{+0.09}_{-0.07}$.  
More details as well as additional
results can be found in the recent salt phase publication.\cite{nsp}

\section*{Acknowledgments}
This research was supported by:  Canada:  Natural Sciences and
Engineering Research Council, Northern Ontario Heritage Fund, Atomic
Energy of Canada, Ltd., Ontario Power Generation, High Performance
Computing Virtual Laboratory, Canada Foundation for Innovation; US:
Dept. of Energy, National Energy Research Scientific Computing Center;
UK: Particle Physics and Astronomy Research Council.

\end{document}